\documentclass[a4paper,11pt]{article}
\usepackage{pos}

\title{Interpolation of rational functions in loop calculations by using $p$-adic numbers}

\author*{Herschel A. Chawdhry}

\affiliation{Department of Physics, Florida State University,\\
77 Chieftan Way, Tallahassee FL 32306, USA}

\emailAdd{hchawdhry@fsu.edu}

\abstract{
The calculation and manipulation of large multi-variable rational functions is a key bottleneck in multi-loop calculations. In these conference proceedings, based on my article~\cite{Chawdhry:2023yyx}, I present a technique to interpolate such rational functions in a compact form by evaluating them at special integer points chosen for their properties under a $p$-adic absolute value. I apply the technique to examples of large rational functions appearing in 2-loop 5-point massless non-planar amplitude calculations in Quantum Chromodynamics (QCD). The number of required numerical probes (per field) is found to be around 25 times smaller than in conventional techniques, and the obtained result is 130 times smaller.}

\FullConference{Loops and Legs in Quantum Field Theory (LL2024)\\
 14-19, April, 2024\\
Wittenberg, Germany\\}

\usepackage{graphicx}% Include figure files
\usepackage{dcolumn}% Align table columns on decimal point
\usepackage{bm}% bold math
\usepackage{amsfonts}
\usepackage{enumitem}% want to avoid column break on page 3 at start of enumerate, using [beginpenalty=10000]

\newcommand{\Q}{\mathbb{Q}}

\newcommand{\F}{\mathbb{F}}
\newcommand{\bigO}{\mathcal{O}}

\begin{document}
\maketitle

\section{Introduction}\label{sec:intro}
The calculation of large rational functions is a central bottleneck in multi-loop amplitude computations, and has indeed been discussed in many talks at this Loops And Legs conference. As will be described shortly, popular approaches for calculating these rational functions include, on the one hand, numerical interpolation methods to bypass large intermediate expressions, and on the other hand, symbolic methods exploiting partial fractioning to produce major simplifications in the size of (intermediate and final) expressions. The subject of this conference proceedings paper, based on my article~\cite{Chawdhry:2023yyx}, is a new technique to interpolate rational functions directly in partial-fractioned form, thereby combining the benefits of the two approaches. The technique uses evaluations at special integer points chosen for their properties under a so-called $p$-adic absolute value. The rational functions are interpolated one partial-fractioned term at a time, exploiting the simplification provided by partial-fractioning and exposing hints of additional patterns and structure that can be exploited in future work.

In symbolic computations involving arithmetic on polynomials or rational functions, the appearance of large intermediate expressions is a ubiquitous problem. Numerical interpolation techniques to bypass this problem have a long history in computer algebra~\cite{Borosh1966ExactSO,collins1968computing,10.1145/321784.321787,10.1145/800206.806398,zurGathenGerhard} and have in recent years been directly adopted with great success by the particle physics community~\cite{vonManteuffel:2014ixa,Peraro:2016wsq,Klappert:2019emp,Peraro:2019svx,Klappert:2020aqs}. They work by replacing symbolic arithmetic with numerical arithmetic in a suitable field, e.g. the (prime) finite fields $\F_p$ or the $p$-adic fields $\Q_p$, and then reproducing the exact symbolic result of a sequence of arithmetic operations by interpolating it from sufficiently many numerical samples. In amplitude calculations, the interpolation itself is typically quick and the computational cost of the approach is largely determined by the time spent performing numerical evaluations, which is in turn determined by the size and complexity of the final result to be interpolated.

In high-energy physics applications of such numerical interpolation techniques, several optimisations have been explored.
The number of probes required can be reduced by a factor of 2 by guessing~\cite{Abreu:2018zmy,Heller:2021qkz} the common denominator of a rational function.
Refs.~\cite{Laurentis:2019bjh,DeLaurentis:2020qle} reconstruct partial-fractioned expressions from very high-precision floating-point evaluations.
Within a finite-field context, some benefits may also be obtained by reconstructing in one variable at a time and performing single-variable partial fractioning at some intermediate stages~\cite{10.1145/800206.806398,Bendle:2019csk,Badger:2021imn,Badger:2021nhg,Badger:2021ega,Abreu:2021asb}, possibly in conjunction with expanding in $\epsilon$, where $D=4-2\epsilon$ is the spacetime dimension variable.
Techniques based on algebraic geometry and evaluations in $\Q_p$ have been proposed~\cite{DeLaurentis:2022otd,DeLaurentis:2023nss,DeLaurentis:2023izi} for eliciting information about the numerator of a rational function prior to performing a finite-field reconstruction, and Ref.~\cite{Campbell:2022qpq} mentions combining these with the methods of Ref.~\cite{Laurentis:2019bjh}. The interpolation itself can sometimes be performed in quasi-linear time by applying the Fast Fourier Transform to evaluations performed at roots of unity~\cite{Chawdhry:2020xlk}.

In parallel with the above numerical techniques, recent years have seen symbolic multiple-variable partial fractioning algorithms employed to simplify the final (and also intermediate) results of heavy calculations, producing simplifications by up to 2 orders of magnitude~\cite{Abreu:2019odu,Boehm:2020ijp,Agarwal:2021grm,Agarwal:2021vdh,Bendle:2021ueg,Heller:2021qkz,Chawdhry:2021mkw,Badger:2022mrb}. An example is shown in Table~\ref{tab:R_simplification_partial_fractioning}, obtained by taking the largest rational function appearing in the integration-by-parts (IBP) table for the 2-loop full-colour QCD amplitudes for $pp \rightarrow \gamma \gamma j$~\cite{Agarwal:2021vdh} and partial fractioning it using the \textsc{MultivariateApart}~\cite{Heller:2021qkz} library. In the present work, this function will be denoted $R_*$ and it will be used as a working example throughout the paper. It can be seen from the table that the partial-fractioned form of $R_*$ is $\bigO(100)$ times smaller than its common-denominator form.

\begin{table}
\begin{center}
\caption{
\label{tab:R_simplification_partial_fractioning}
Simplification of $R_*$ under partial~fractioning.
Common-denominator form has numerator fully expanded and denominator fully factorised.
Partial-fractioned form is obtained using \textsc{MultivariateApart}~\cite{Heller:2021qkz} with option \texttt{UseFormProgram->True}.
(See Table~\ref{tab:results} for results obtained in this work by numerical interpolation directly in partial-fractioned form.)
Sizes are as reported using \texttt{ByteCount} in \textsc{Mathematica}.
Number of free parameters is obtained by counting the number of terms in the fully-expanded numerator(s). 
}
\begin{tabular}{c|c|c}
Form of expression	&	Size	& Parameters to fit	\\
\hline
Common-denominator	& 605 MB &	1,369,559	\\
Partial-fractioned	& 4 MB	& 14,558
\end{tabular}
\end{center}
\end{table}

Although it is well known that the denominator of the common-denominator form of IBP expressions like $R_*$ usually factorise into simple polynomial factors, this alone does not explain the simplification in Table~\ref{tab:R_simplification_partial_fractioning}.
To see this, several rational functions were considered, taken from the solutions to the 2-loop 5-point massless non-planar IBP equations.
For each rational function $R$, a second rational function $\tilde{R}$ was constructed by taking $R$ in common-denominator form and replacing with random numbers all coefficients in its fully-expanded numerator, while leaving the denominator unchanged.
It was observed that each $R$ simplifies upon partial fractioning, and the simplification factor is largest for the largest rational functions.
Yet upon partial fractioning $\tilde{R}$, no simplification occurs; indeed the partial-fractioned form of $\tilde{R}$ is typically slightly larger than its common-denominator form, regardless of whether it is measured using \texttt{ByteCount} or the number of free numerator parameters.
It can therefore be concluded that the above-mentioned simplification of $R_*$ upon partial fractioning does not occur for generic rational functions, but is instead a special property of $R_*$, which is conjectured here to generalise to many IBP and amplitude expressions.\footnote{We emphasise that the selection of $R_*$ as working example was not on the basis of any such properties, but was on the contrary because it is an exceedingly complicated expression that is on the boundary of current computational techniques.}

In this work, a new technique is presented to interpolate rational functions directly in partial-fractioned form, to improve the speed (and hence reach) of loop calculations.
The technique uses $p$-adic probes to reconstruct the rational functions one partial-fractioned term at a time, giving a powerful capability to better identify, understand, and exploit the structures in these functions.
Indeed, it will be seen in sec.~\ref{sec:results} that the technique requires 25 times fewer numeric ($\Q_p$) probes than conventional ($\F_p$-based) reconstruction, and leads to a 130-fold reduction in the size of the final result.
The results furthermore reveal hints of further patterns and it is therefore expected that the technique will prove to be a valuable tool to study and exploit them.

To understand the reason for the simplification in Table~\ref{tab:R_simplification_partial_fractioning} and guide a strategy for exploiting it, the program \textsc{MultivariateApart}~\cite{Heller:2021qkz} was applied to several examples of $R$ and $\tilde{R}$. In each case the resulting expressions
\begin{equation}\label{eq:R_pfed}
R = \sum_i \frac{n_i}{d_i},
\end{equation}
\begin{equation}\label{eq:Rtilde_pfed}
\tilde{R} = \sum_j \frac{\tilde{n}_j}{d_j}
\end{equation}
were compared.
For all the examples studied, it was observed that the sum in eq.~\eqref{eq:R_pfed} contains fewer terms than the sum in eq.~\eqref{eq:Rtilde_pfed}.
Furthermore, all of the terms in eq.~\eqref{eq:R_pfed} also appear in eq.~\eqref{eq:Rtilde_pfed}, albeit with different numerators---in other words, $\{d_i\}$ is a subset of $\{d_j\}$.
It was furthermore noted that if the partial-fractioned terms that are present in $\tilde{R}$ but vanish in $R$ could be identified in advance, it would give a large simplification. In the case of $R_*$, it was estimated that this simplification would be a factor of 28 compared to the common-denominator form, reducing the number of free parameters from 1,369,559 to 48,512.
For this reason, the core aim of the method presented in sec.~\ref{sec:method} is to identify, as cheaply as possible, which partial-fractioned terms vanish.
The remaining factor of $\frac{48,512}{14,558} \approx 3.3$ between this and the figure in Table~\ref{tab:R_simplification_partial_fractioning} arises because many of the partial-fractioned terms in eq.~\eqref{eq:R_pfed} have numerators containing fewer terms than the most generic polynomial that could be expected; the exploitation of this further simplification is left to future work. In addition, as already mentioned, the results (see sec.~\ref{sec:results}) suggest that further patterns and structure are present, which could be explored and exploited in future work to obtain significant further speed-ups.

\section{Method}\label{sec:method}
In order to exploit the observations from sec.~\ref{sec:intro}, a method was devised to reconstruct rational functions directly in partial-fractioned form eq.~\eqref{eq:R_pfed}, one partial-fractioned term $\frac{n_i}{d_i}$ at a time.
A set of all possible denominators $\{d_i\}$ is straight-forward to determine by examining the easily-obtainable 
denominator~\cite{Abreu:2018zmy,Heller:2021qkz} of the common-denominator-form expression.
As explained in sec.~\ref{sec:intro}, the speedup in this paper will arise because for many of the possible denominators $d_i$, the corresponding $n_i$ is zero.
Reconstructing one partial-fractioned term at a time ensures that if a partial-fractioned term vanishes, one can notice this cheaply and avoid reconstructing its numerator.
A key further advantage of reconstructing one partial-fractioned term at a time is that our method will scale well for even larger rational functions than $R_*$, because it allows interpolation to be performed without needing to invert large systems of linear equations.

Reconstructing one partial-fractioned term at a time also has other benefits, which are foreseen here but will be left to further work: for instance, noting that the bottleneck in cutting-edge calculations is sometimes a very small number of particularly large rational functions, it can be expected that reconstructing one partial-fractioned term at a time would give maximum scope for on-the-fly observation of patterns that can be exploited in the remaining partial-fractioned terms. Examples of this might be the optimal choice of numerator variables for particular combinations of denominator factors, or the appearance of commonly-occurring integer or polynomial prefactors in the numerators of some partial-fractioned terms, or even (as is observed post-hoc in sec.~\ref{sec:results}) the appearance of identical numerators in several partial-fractioned terms.
Additionally, this method of reconstructing one partial-fractioned term at a time can provide a powerful tool to better analytically understand---and eventually further exploit---the simplification that partial fractioning produces for rational functions in amplitudes and IBP expressions.

The method uses evaluations in the $p$-adic numbers $\Q_p$, which have been studied by mathematicians for over a century and can be applied to a variety of areas of physics~\cite{Dragovich:2009hd} ranging from $p$-adic quantum mechanics~\cite{Djordjevic:1999vi} to quantum field theory~\cite{Smirnov:1990uz,Smirnov:1992sr} to string amplitudes~\cite{Volovich:1987zc, Freund:1987ck,Marinari:1987tn} to, as pointed out in~\cite{DeLaurentis:2022otd}, the study of the singular limits of polynomials and rational functions,
which is helpful for studying the partial-fractioning of
rational functions as desired in this work. The reader is referred to my main article~\cite{Chawdhry:2023yyx} and references therein for more information on the topic, an introduction to common notation and notions such as the $p$-adic absolute value $|\cdot|_p$, and a summary of the properties of $\Q_p$ that this interpolation method relies upon.

Let $R$ denote the rational function we wish to reconstruct and let $N$ be the number of variables it contains. We start by observing, as a consequence of eq.~\eqref{eq:R_pfed}, that if we can find a special $p$-adic point $\bar{x}$ at which the denominator $d_k$ of one partial-fractioned term $n_k / d_k$ becomes $p$-adically smaller than all the others, i.e. if for some $\bar{x} \in \Q_p^N$,
\begin{equation}\label{eq:pick_out_unique}
\exists k : \forall i \neq k, \quad |d_k (\bar{x})|_p < |d_i (\bar{x})|_p,
\end{equation}
then evaluating the complete rational function $R$ at that $p$-adic point $\bar{x}$ will give a series
\begin{equation}\label{eq:R_x}
R(\bar{x}) = \frac{n_k(\bar{x})}{d_k(\bar{x})} + \bigO(p^{-m+1}),
\end{equation}
where $m = -\log_p\left( |d_k(\bar{x})|_p \right)$.\footnote{In this work, $\log_p$ does not denote the $p$-adic logarithm sometimes seen in the mathematical literature, but instead just an ordinary logarithm with base $p$.}
Eq.~\eqref{eq:R_x} is a consequence of eq.~\eqref{eq:R_pfed}, but obviously the result of a numerical evaluation does not depend on whether it is performed using the partial-fractioned form of $R$, or its common-denominator form, or even a ``black-box'' program that produces numerical evaluations of $R$ without knowledge of its explicit symbolic form.
In general, the series~\eqref{eq:R_x} is $\bigO(p^{-m})$ and the coefficient of $p^{-m}$ gives useful information about $n_k(\bar{x})$.
In particular, if $n_k = 0$, the $\bigO(p^{-m})$ term will vanish and so the leading term of the series $R(\bar{x})$ will be $\bigO(p^{-m+1})$ instead.
Furthermore, even when $n_k \neq 0$, we can use eq.~\eqref{eq:R_x} to obtain the leading $p$-adic digit of $n_k(\bar{x})$, in effect obtaining a finite-field evaluation of $n_k$.
By repeating for other values of $\bar{x}$ that still satisfy eq.~\eqref{eq:pick_out_unique} for the same $k$, we can gather sufficient information to reconstruct the analytic form of $n_k$ modulo $p$. This procedure can then be repeated for other fields $\Q_p$ in order to then obtain the complete expression for $n_k$ using the Chinese remainder theorem. It is vital to perform this last step \emph{before} proceeding to probe or reconstruct other partial-fractioned terms.
For a more complete description of this procedure, the reader is referred to the main article~\cite{Chawdhry:2023yyx}.

There are many important details that are beyond the scope of this brief proceedings paper but that are essential for the implementation of this interpolation technique. For example, the choice of the special points $\bar{x}$ in eqs.~\eqref{eq:pick_out_unique} and~\eqref{eq:R_x} merits elaboration: while it is straight-forward to see that certain choices of $p$-adic point might pick out subsets of the partial-fractioned terms in eq.~\eqref{eq:R_pfed}, picking out a single partial-fractioned term at a time is non-trivial. The procedure by which this can be done is explained comprehensively in my main article~\cite{Chawdhry:2023yyx}, along with other details and explanations that have been omitted from this brief proceedings paper.

\section{Results and discussion}\label{sec:results}

By employing the technique presented in sec.~\ref{sec:method}, the rational function $R_*$ was reconstructed in full. This was achieved using no knowledge of $R_*$ other than its mass dimension, its common denominator, and the results of ``black-box'' probes at integer-valued kinematic points.
As shown in Table~\ref{tab:results}, the reconstructed result is 130 times smaller than the common-denominator form targeted by conventional techniques.
The reconstruction required $\bigO(6 \times 10^4)$ numerical ($\Q_p$) evaluations per prime, whereas the conventional ($\F_p$-based) approaches would require $\bigO(1.4 \times 10^6)$ probes per prime. Furthermore, the parameters in the partial-fractioned form are generally simpler than those in common-denominator form, and so for most parameters only 3 or 4 $p$-adic fields were required (plus one for checks), with $p \sim \bigO(100)$.

\begin{table}
\begin{center}
\caption{
\label{tab:results}
Comparison of original and reconstructed form of $R_*$.
Original expression is in common-denominator form, with numerator fully expanded and denominator fully factorised.
Sizes are as reported using \texttt{ByteCount} in \textsc{Mathematica}.
Number of free parameters is obtained by counting the number of terms in the fully-expanded numerator(s). 
}
\begin{tabular}{c|c|c}
Expression	&	Size	& Parameters to fit \\
\hline
Original & 605 MB &	1,369,559 \\
Reconstructed	& 4.5 MB	& 52,527 (\emph{of which 15,403 non-zero})
\end{tabular}
\end{center}
\end{table}

The reconstructed result exhibits further structure which in future work it would be beneficial to study and exploit.
This is best seen and discussed with an example.
We will consider the following reconstructed terms
\begin{multline}\label{eq:some_reconstructed_terms}
\frac{\frac{45}{1024} s_{45}^6 s_{12}^3}{(D-3) s_{34}^4 s_{51} (-s_{23}+s_{45}+s_{51})^3}
+\frac{\frac{9}{5120} s_{45}^6 s_{12}^3}{(D-1) s_{34}^4 s_{51} (-s_{23}+s_{45}+s_{51})^3}\\
-\frac{\frac{693}{5120} s_{45}^6 s_{12}^3}{(2 D-7) s_{34}^4 s_{51} (-s_{23}+s_{45}+s_{51})^3}
-\frac{\frac{3}{1024} s_{45}^6 s_{12}^3}{s_{34}^4 s_{51} (-s_{23}+s_{45}+s_{51})^3}\\
+\frac{-\frac{45 s_{45}^6 s_{51}^2}{1024}-\frac{135 s_{45}^6 s_{51} s_{12}}{1024}-\frac{135 s_{45}^6 s_{12}^2}{1024}}{(D-3) s_{34}^4 (s_{23}-s_{45}-s_{51})^3}
+\frac{-\frac{9 s_{45}^6 s_{51}^2}{5120}-\frac{27 s_{45}^6 s_{51} s_{12}}{5120}-\frac{27 s_{45}^6 s_{12}^2}{5120}}{(D-1) s_{34}^4 (s_{23}-s_{45}-s_{51})^3}\\
+\frac{\frac{693 s_{45}^6 s_{51}^2}{5120}+\frac{2079 s_{45}^6 s_{51} s_{12}}{5120}+\frac{2079 s_{45}^6 s_{12}^2}{5120}}{(2 D-7) s_{34}^4 (s_{23}-s_{45}-s_{51})^3}
+\frac{-\frac{3 s_{45}^6 s_{51}^2}{1024}-\frac{9 s_{45}^6 s_{51} s_{12}}{1024}-\frac{9 s_{45}^6 s_{12}^2}{1024}}{s_{34}^4 (-s_{23}+s_{45}+s_{51})^3},
\end{multline}
which form a small part of our full reconstructed result.
Here $s_{ij}$ are the 5 kinematic variables of $R$.

Firstly, it should be mentioned that 70\% of the free parameters that were fitted turned out to be zero, as anticipated from the discussion at the end of sec.~\ref{sec:intro}.
This is can be seen in expression~\eqref{eq:some_reconstructed_terms} in the following way. Expression~\eqref{eq:some_reconstructed_terms} contains 16 numerator terms and therefore accounts for 16 of the 15,403 non-zero free parameters mentioned in Table~\ref{tab:results}.
Considering the first term in~\eqref{eq:some_reconstructed_terms}, we note that a priori there was no reason for the numerator to only contain a term $\sim s_{45}^6 s_{12}^3$; it could equally well have contained other mass-squared-dimension-9 combinations of $s_{45}$ and $s_{12}$, such as $s_{45}^2 s_{12}^7$.
To obtain~\eqref{eq:some_reconstructed_terms} a total of 220 free parameters were therefore fitted, of which 204 turned out to be zero.
Expression~\eqref{eq:some_reconstructed_terms} thus accounts for 220 of the 52,527 free parameters mentioned in Table~\ref{tab:results}.
Identifying the vanishing parameters in advance would reduce the number of parameters to be fitted from 52,527 to 15,403, and reduce the number of probes correspondingly.

Secondly, some of the numerators in our reconstructed result are linearly related to each other by a simple integer multiple.
Looking at the 2 numerators on the 3\textsuperscript{rd} line of expression~\eqref{eq:some_reconstructed_terms}
\begin{equation}
n_1 = - \frac{45 s_{45}^6 s_{51}^2}{1024}-\frac{135 s_{45}^6 s_{51} s_{12}}{1024}-\frac{135 s_{45}^6 s_{12}^2}{1024},
\end{equation}
\begin{equation}
n_2 = - \frac{9 s_{45}^6 s_{51}^2}{5120}-\frac{27 s_{45}^6 s_{51} s_{12}}{5120}-\frac{27 s_{45}^6 s_{12}^2}{5120},
\end{equation}
it can be noticed that $n_1 = 25 n_2$.
If such relations can economically be discovered prior to reconstruction, it would further reduce the number of free parameters to be fitted and thus the number of probes required.

Thirdly, it was noticed that in some cases it is possible to combine several of our reconstructed terms and obtain a simpler expression. For example, if we combine together all the terms in expression~\eqref{eq:some_reconstructed_terms}, we obtain the following simple term:
\begin{equation}\label{eq:reconstructed_terms_combined}
-\frac{\frac{3}{512} D \left(D^2-4\right) s_{45}^6 (s_{51}+s_{12})^3}{(D-3) (D-1) (2 D-7) s_{34}^4 s_{51} (-s_{23}+s_{45}+s_{51})^3}.
\end{equation}
Note however that the first two properties do not necessarily imply the third, and it was observed from examining other reconstructed terms that combining them in this manner does not always simplify them.
The results in Table~\ref{tab:results} do not employ any such recombination of terms, and further study is required to understand which cases are amenable to such simplification and to devise a manner to exploit it during the reconstruction itself, rather than afterwards.
This is an interesting direction for exploration, with the potential to yield a further order-of-magnitude reduction in the number of free parameters to be fitted, the number of probes required per prime field, and the size of the final result.
Additionally, since the numerical coefficient $\frac{3}{512}$ in~\eqref{eq:reconstructed_terms_combined} is somewhat simpler than coefficients like $\frac{2079}{5020}$ in~\eqref{eq:some_reconstructed_terms}, fewer prime fields would be required to fit this coefficient.

It is worthwhile to note that although the patterns and structure exploited in this work---as well as those left for future work---could be studied post-hoc by partial-fractioning an expression obtained by conventional means, this work's technique of reconstructing one partial-fractioned term at a time provides the capability to study and exploit these structures \emph{during} the reconstruction.
For cutting-edge calculations where obtaining any analytic expression in the first place is the principal challenge and goal, this new capability can be a valuable asset.

Going further, emphasis should be placed on the desirability of analytically studying the simplifications explored in this work, possibly in conjunction with the observations in Refs.~\cite{Laurentis:2019bjh,DeLaurentis:2022otd,Campbell:2022qpq}.
It is hoped that the techniques presented in this work will prove to be useful tools in this regard, with benefits for our theoretical understanding as well as the speed of calculations.

\section{Conclusion and outlook}\label{sec:conclusion}
In this proceedings paper, based on my article~\cite{Chawdhry:2023yyx}, a new interpolation technique was presented to reconstruct rational functions directly in partial-fractioned form.
It uses $p$-adic evaluations to harness the major simplification of rational functions under partial fractioning.
It was shown that this simplification does not occur for more generic rational functions, and so it instead appears to be a specific feature of the rational functions appearing in loop calculations.

The interpolation technique was demonstrated using the example of $R_*$, the largest rational function in one of the largest IBP coefficients needed for any 2-loop 5-point massless non-planar QCD amplitude. It was found that the technique can reconstruct such functions using 25 times fewer numerical ($\Q_p$) probes than conventional techniques based on finite-field probes, and yields a 100-fold simplification in the size of the reconstructed result.

Having demonstrated this technique on the large rational functions appearing in massless 2-loop 5-point QCD calculations, the natural next step would be to apply it to 2-loop 5-point processes with masses, most of which currently remain unknown.
Preliminary work indicates that the technique generalises straight-forwardly to such processes. More generally, experience shows that partial-fractioning produces simplifications in amplitudes whenever several kinematic scales are present, and so the technique is expected to be applicable to a wide range of higher-point or higher-loop amplitudes.

A number of technical improvements could be implemented to further extend the improvements obtained with this method. For example, one could recycle the probes used during reconstruction, which we expect will give a significant further reduction in the overall number of probes required. In addition, since this work focussed on reducing the number of probes while remaining agnostic as to the choice of computational implementation of $p$-adic numbers, it would be useful in future work to explore various implementation strategies and compare their costs relative to each other and relative to conventional $\F_p$ probes. At present it would be prudent to assume  $p$-adic probes to be slower than $\F_p$ probes, but it should also be highlighted that using small-valued primes $p \sim \bigO(100)$, as in this work, is likely to be beneficial, especially for filtering out partial-fractioned terms that vanish, regardless of the way in which one implements $p$-adic numbers on a computer.

Finally, it was observed that the reconstructed result for $R_*$ displays further patterns and structures which would be worthwhile to study, understand, and exploit in future work.
These observations provide hints of the potential to obtain even further improvements in the speed and reach of this calculational method, as well as potential avenues for starting to seek further understanding of the physical origin of these simplifications and of the structure of the rational functions appearing in scattering amplitudes and IBPs.

\begin{acknowledgments}
We are grateful to Federico Buccioni, Fabrizio Caola, Fernando Febres Cordero, Stephen Jones, Stefano Laporta, Andrew M\textsuperscript{c}Clement, and Alex Mitov for helpful discussions. We also thank the authors of Ref.~\cite{Agarwal:2021vdh} for providing analytic expressions for the IBP coefficients used as examples in this work.

This work has been funded by the European Research Council (ERC) under the European Union's Horizon 2020 research and innovation programme (grant agreement no. 804394 \textsc{hipQCD}).
The work was also supported in part by the U.S. Department of Energy under grant DE-SC0010102.

\end{acknowledgments}

\bibliographystyle{JHEP}
\bibliography{p-adic_reconstruction_loopsAndLegs2024_proceeding.bib}

\end{document}